# Estimating the Effects of Syrian Civil War[*]

Aleksandar Kešeljević        Rok Spruk

## Abstract

*We examine the effect of civil war in Syria on economic growth, human development and institutional quality. Building on the synthetic control method, we estimate the missing counterfactual scenario in the hypothetical absence of the armed conflict that led to unprecedented humanitarian crisis and population displacement in modern history. By matching Syria's growth and development trajectories with the characteristics of the donor pool of 66 countries with no armed internal conflict in the period 1996-2021, we estimate a series of growth and development gaps attributed to civil war. Syrian civil war appears to have had a temporary negative effect on the trajectory of economic growth that almost disappeared before the onset of COVID19 pandemic. By contrast, the civil war led to unprecedented losses in human development, rising infant mortality and rampantly deteriorating institutional quality. Down to the present day, each year of the conflict led to 5,700 additional under-five child deaths with permanently derailed negative effect on longevity. The civil war led to unprecedent and permanent deterioration in institutional quality indicated by pervasive weakening of the rule of law and deleterious impacts on government effectiveness, civil liberties and widespread escalation of corruption. The estimated effects survive a battery of placebo checks.*

---

[*] Kešeljević: Professor of Economics, School of Economics and Business, University of Ljubljana, Kardeljeva ploščad 17, SI-1000 Ljubljana. E: saso.keseljevic@ef.uni-lj.si. Spruk: Assistant Professor of Economics, School of Economics and Business, University of Ljubljana, Kardeljeva ploščad 17, SI-1000 Ljubljana. E: rok.spruk@ef.uni-lj.si.

# I  Introduction

Since the time immemorial, the economic history has taught us that civil wars cause very large economic, social and environmental costs for the society. Compared to international wars, civil wars are oftentimes more devastating due to large-scale displacement of population, severe causalities, and millions of refugees (Collier 1999). After long periods of war, the economies have a tendency to recover rapidly whilst the tendency of decline is strong in wars of short duration. Without the loss of generality, the notion that civil war poses a development in reverse and entails severe economic losses and high opportunity costs of conflict is not an overstatement. The economic effects of civil war are traditionally examined through an ex-post evaluation of the costs of armed conflict. The analysis may be carried out at the level of countries, regions or cities that allows us to capture different dimensions of effects.

In this paper, we examine the economic cost of the Syrian civil war that began to unfold in 2011 after a violent crackdown of public protests during the Arab Spring by the Al Assad regime. The toll of Syrian civil war unveils an unprecedented humanitarian and economic loss of the armed conflict between Al Assad regime and Syrian opposition forces. The Syrian civil war led to one of the largest outflows and displacement of population. The civil war drained the population to an unprecedented degree in modern history with estimated 13 million displaced people requiring humanitarian assistance, out of which around five million crossed the borders, leading to one of the most severe refugee crises in history. Although the existing estimates provide a clear insight into the humanitarian costs of war, broader economic, social and institutional quality impacts of the Syrian civil war await further analysis and quantification. In particular, we estimate the causal effect of Syrian civil war by building the missing counterfactual scenario of growth and development trajectories in the hypothetical absence of war through multiple comparisons of Syria with a group of countries that have not undergone any armed conflict in the period of our investigation between 1996 and 2021 using the counterfactual outcome modelling approach.

In the extensive literature on the evaluation of the effects of civil war, the notion of counterfactual scenario has become widely accepted in recent years. Under this approach, the economic and social effect of civil war is evaluated by estimating the missing counterfactual scenario of growth and development in the hypothetical absence of the war by making use of the synthetic control method originally proposed by Abadie and Gardeazabal (2003) and further extended by Abadie et. al. (2015) and many others (Gilchrist et. al. 2022). The general thrust behind the application of synthetic control method to evaluate the effect of civil wars is intuitive. By way of comparison of the war-affected areas with a plausible control sample of areas that did not undergo armed internal conflict, growth and development effects of civil war can be estimated. In terms of further example, Bluszcz and Valente (2022) employ the synthetic control method to estimate the effect of Donbass war on Ukraine and its most affected territories in the Donbas basin, and find evidence of particularly large and devastating effects on both Ukraine and the breakaway regions in the eastern part of the country relative to their control groups where no armed conflict took place. Furthermore, Echevarría and García-Enriquez (2019) exploit the synthetic control method to estimate the effect of Arab Spring in Libya in 2011 that led to a full-fledged civil war on Libyan economy, and find evidence of large cumulative losses of GDP in both absolute and per capita terms. Similar analyses using synthetic control method have been carried out to evaluate the effect of Second Intifada on Israeli economy (Horiuchi and Mayerson 2015), the effect of Kurdish separatism on the economic growth trajectory of Turkey (Bilgel and Can Karahasan 2017, 2019), the effect of Arab Spring on Egyptian economy (Echevarría and García-Enriquez 2020), and the effect of Yugoslav civil war on the economies of ex-Yugoslav republics (Kešeljević and Spruk 2021) among several others (Bove et. al. 2017).

Although synthetic control method has not been untouched by scholarly criticism of its distributional assumption and strength of inference, (Hsiao et. al. 2012, Gardeazabal and Vega-Bayo 2016, Li and Bell 2017, Wan et. al. 2018, Gharehgozli 2021), we emphasize several noteworthy advantages of the synthetic control method in estimating the effects of civil war compared to alternative estimators. In examining the costs of armed conflict, three major approach streams have been proposed. First, compared to cost-accounting approach favoured by Arunatilake et. al. (2001), Bilmes and Stiglitz (2006), Skaperdas et. al. (2009) and several others, synthetic control method does not necessitate multiple calculations of a broad variety conflict costs. The general thrust of cost-accounting calculations is the innate reliance on the data provided by governments and international agencies which has been subject to criticism since the quality and reliability may be questionable. In addition, cost-accounting approach is also prone to double counting of costs. By contrast, synthetic control method allows us to perform statistical inference for each outcome directly without double counting of the costs. Second, the costs of conflict may be examined through the application of both static and dynamic panel-data as well as basic and advanced time-series analysis of the cost of conflict favoured by Enders et. al. (1992) and Barro and Lee (1994) among several other scholars. However, the these estimators are generally not robust to the presence of structural confounding variables and the multiple sources of unobserved heterogeneity that co-shape and simultaneously influence the trajectories of economic growth and development. The standard approach to deal with confounding issues using panel-data approach is to add a battery of additive and multiplicative fixed effects that capture country-specific variations in geo-political alliances and networks that may over-absorb the effect of civil war in difference-in-differences model of conflict as the preferred estimator in the evaluation of conflict costs. By contrast, synthetic control analysis is able to generalize difference-in-differences analysis and does not hinge on parallel trend assumption as a prerequisite to estimate the effect of civil war, and also embeds time trends, unobserved effects and time-varying heterogeneity directly into the factor model used to estimate the counterfactual scenario of the civil war. In a similar vein, the synthetic control estimator performs well in small samples (Ferman and Pinto 2021), it does not necessitate the arbitrary search for specification (Gilchrist et. al. 2022) and avoids the extrapolation outside the range of data (Abadie 2021). Therefore, by applying the synthetic control estimator to study the effects of civil war on Syrian economy and society, plausible and reasonable estimates of the causal effect of war can be estimated.

The rest of the paper is structured as follows. Section II reviews the related literature. Section III presents our identification strategy. Section IV presents the data along with treatment and control samples. Section V discusses the results and their validity. Section VI concludes.

## II     Related Literature

Public protests and demonstrations by the Syrian opposition began after the Arab spring protests in Egypt and Tunisia in 2011 demanding an end of corruption and abuses of the security forces. What began as peaceful protests escalated into a full-scale armed conflict among many parties with increasingly sectarian character and supported by foreign forces (Iran, Turkey, USA, Russia). Since the beginning the war, Syria remains deeply fragmented and divided into three zones. The Syrian government controls much of the country and all of its major cities by support from Iran and Russia. Northeast is in the hands of a Kurdish force backed by the U.S until their withdrawal. The northwest is under Turkish control with a mix of opposition forces.

Civil war in Syria has caused enormous suffering to the civilians and society with damaging effects due to displacement of people, thousands of dead, millions of refugees and large-scale social and economic costs. The war set the country back decades in economic, social and human development. A lot of civil war's damage is irreversible, due to inhibited growth potentials and undermined trust among the main ethnic and religious groups and by leaving a deep mark on the people due their tragedies and human suffering on a daily basis.

The economic theory of war suggests that wars cause different kinds of (in)direct costs to the society. Firstly, war causes destruction which results in a GDP decline. Secondly, war results in loss of human capital. Thirdly, war increases fiscal deficits and crowds out public expenditures on education and health. Fourthly, civil war inhibits economic activity due to declining capital stock. And lastly, neighboring countries are affected by the spillover effects through a multitude of impacts of refugee flows.

*GDP decline.* The existing literature on the economic effects of war shows that regardless of which approach and model we use, the impact of civil war on output is always disastrous and implies large economic losses (Mueller, 2013; Collier, 1999; Gates et al., 2012; Kang and Meernik, 2005). The report by World Bank (2017) shows that Syria's GDP contracted by 63 percent between 2011 and 2015 compared with 2010. In terms of further example, Giovannetti and Perra (2019) estimates Syria's GDP 43% fall in 2013 through non-conventional approach by using satellite data (i.e. night light intensity) to calculate the average rate of decrease in luminosity to approximate the GDP decline. A relatively meaningful insight into the costs of war is a comparison of the development trajectories between war zones and areas not affected by the civil war (Collier and Hoeffler 2002; Cerra and Saxena 2008, Kešeljevič and Spruk 2021). World Bank's (2017) estimation of differences between counterfactual and actual GDP between 2011 and 2016 shows that the cumulative loss in GDP is $226 billion in 2010 prices, about four times the 2010 GDP in Syria. Disproportionately large decline has been in agriculture (WB, 2017), manufacturing (Gobat and Kostial 2016), finance (Samir 2020), housing and energy sector (World Bank 2017a).

*Human capital losses.* Civil wars cause a radical loss of the human capital stock due to psychological traumas (Bratti et al., 2016; Murthy and Lakshiminarayana 2006), educational losses and permanently lower productivity (Lai and Thyne 2007; Eder 2014; Devakumar et al., 2014; Ichino and Winter-Ebmer 2004), displacement of people and human capital outflow (Mueller 2013). According to the UNHRC [2] more than 350,000 people have been killed with 6.2 million displaced people, including 2.5 million children within Syria. In addition, more than 700,000 Syrian nationals are estimated to have sought political asylum in Europe in 2015 and 2016 (Eurostat 2022; World Bank 2017). Out of a pre-conflict population of over 20 million declined to 18,2 mio. 11.9 million people have been forcibly displaced within Syria and across its borders, out of which 5.7 million refugees and 6.2 million internally displaced). Furthermore, A report by Syrian Center for Policy Research (2015) shows that 80% of the population has plunged into poverty compared to 12.4% in 2007.

*Fiscal adjustments and crowding out effect.* Civil war crowds out private investment (Stiglitz and Bilmes 2012) and leads in general to massive diversion of public funds from expenditures that promote education and public health (Collier 1999; Ghobarah et al. 2003; Mueller 2013). The 2017 report by World Bank shows that a whole generation of children in Syria has received inadequate education in. Families are struggling to meet their basic needs and are increasingly reliant on negative practices as putting children out of school to work, entering daughters into early wedlock and allowing children to become involved with armed groups (Gobat and Kostial 2016). Higher war costs and declining fiscal revenues, from 23 percent of GDP in 2010 to less than 3 percent in 2015, led to a fiscal deficit of over 20 percent of the GDP. Civil war and trade embargo reduced Syrian exports by 92% between 2011 and 2015 and the current account deficit reached 28% GDP in 2016 up from 0.65% GDP in 2010 (World Bank 2017).

---

[2] United Nations Human Rights Council

***Declining capital stock.*** Reduced connectivity, higher transportation costs, disruptions in supply chains and networks, increased rent seeking, and erosion of trust largely inhibit investment activity in Syria. Bridges, water resources, roads and other significant assets have become targets and their destruction has precipitated the collapse of economic activities. Damage assessment in three selected cities (i.e. Allepo, Hama, Idlib) showed that the highest tangible damage is in the housing, energy and health sector (World Bank 2017a). Outbreak of violence inhibits economic activity due to falling investments and reduces the incentives to pursue productive activities. Even by replacing the capital stock by itself it would not be sufficient to bring the Syrian economy back to its pre-conflict level without rebuilding institutions and networks (Ford 2019; Gobat and Kostial 2016).

***Spillover effects.*** War-related externalities exist due to the negative effect on other countries by destroying infrastructure and capital close to the border (Murdoch and Sandler 2002), decrease of trade volume (Martin et al. 2008), acceptance of refugees and reputational loss (Brück and Henning 2016). Civil war in Syria has caused refugee shock waves through many neighboring countries. According to the United Nations High Commissioner for Refugees (2022) the total number of Syrians refugees in Lebanon, Turkey, Jordan, Iraq, Egypt and North Africa is 5.7 million. However these number do not include 0.4 to 1.1 million unregistered refugees . The refugee problem has also other consequences on the hosting countries. Syrian refugee inflows led to small but statistically significant informal employment losses and labor cost advantages in Turkey (Tumen 2016). ISIS terrorist activity blocked transport routes, increased investment uncertainty and poverty in Kurdistan region of Iraq (World Bank 2015). Due to civil war in Siria from 2011 onward cumulatively GDP reductions correspond to 11.3% of the pre-conflict (2010) GDPs across the Mashreq region (Iraq, Jordan, Lebanon) as shown in the report by International Bank for Reconstruction and Development (2020). In terms of further example, Akugündüz et. al. (2018) analyse how the inflow of Syrian refugees in Turkey affected firm entry and performance, and show that hosting refugees is favorable for firms although firm entry rates do not seem to be affected.

Certain effects of Syrian civil war are barely measurable due to many indirect costs. Several elements in the calculation are missing due to difficulties in assessing them. Some of political, social, security-related, and institutional effects are not easily quantifiable. In addition to damage assessment reported in the existing literature, the costs of war can also be indirectly assessed through the counterfactual analysis. Relatively meaningful insight into the costs of war is comparision of the development trajectories between war zones and areas that were not affected by the war. Literature overview shows two such studies for Syria: The abovementioned report by World Bank (2017) estimates the counterfactual scenario of war for Syria's GDP but only up until 2016. On the other hand, the study by Onder et. al. (2020) attempt to isolate the impact of the Syrian armed conflict and examine the effects on GDP, trade flows, poverty rates and debt burden for Iraq, Jordan and Lebanon but do not carry out a similar analysis for Syria.

## III        Identification Strategy

*III.A Setup*

The goal of our identification strategy is to examine the growth and development cost of the Syrian civil war consistently. To this end, our aim is to estimate the missing counterfactual scenario and simulate the trajectories of economic growth and development in the hypothetical absence of the war. In doing so, we apply the synthetic control estimator (Abadie and Gardeazabal 2003, Abadie et. al. 2015, Costalli et. al. 2017, Echevarría and García-Enríquez 2019, 2020, Gilchrist et. al. 2022) to estimate the causal effect of civil war on Syrian economy and society. Through the application of the potential outcomes framework (Rubin 1973), we are able to simulate the growth and development trajectories of Syria through the combination of attributes of other countries that have similar characteristics but have not undergone an armed conflict in the respective period of investigation. This approach allows us to construct a counterfactual representation of Syria using weighted averages reflecting the resemblances of outcome- and covariate-related characteristics before the outbreak of the civil war.

It should be noted that the causal effect of the Syrian civil war is reflected by the outcome difference between Syria after the civil war and the estimated counterfactual without the armed conflict. The synthetic control group that best reproduces the growth and development attributes of Syria prior to the war allows us to predict the plausible level of outcome if the war never broke out. Provided that the synthetic control group is not tainted by the presence of armed conflict or insurgencies, the outcome difference plausibly reflects the causal effect of civil war on Syrian economy and society.

Suppose that we observe $J + 1$ countries over $t = 1,2, \ldots T$ period where $J$ denotes the country affected by the civil war and $\{2, \ldots J + 1\}$ represent the donor pool of countries not affected by the war. As a mimic of the treatment, civil war occurs at time $T_0$ and lasts in the post-treatment period so that $t < T_0 < T$. Our aim is to measure the impact of the civil war on the set of growth and development outcomes. Without the loss of generality, let $Y_{i,t}^N$ denote the outcome for country $i$ at time $t$ in the absence of the civil war. By contrast, let $Y_{i,t}^I$ be the outcome for country $i$ that would be observed if it were exposed to the civil war. By assuming that the civil war has no effect on the outcome of interest in the pre-war period, it follows that $Y_{i,t}^N = Y_{i,t}^I$ for all $i$ and $t < T_0 + 1$ and we further assume that the civil war only affects Syria. Since the classical estimation and inference procedures for synthetic control method do not explicitly allow for the existence of spillover effects (Cao and Dowd 2019, Di Stefano and Mellace 2020), we do not include the neighboring states[3] into the donor pool of countries unaffected by the civil war. For each $t > T_0$, the effect of the civil war on growth and development outcomes can be written as follows:

$$\theta_{1,t} = Y_{1,t}^I - Y_{1,t}^N = \underbrace{Y_{1,t}}_{\text{observed}} - \underbrace{Y_{1,t}^N}_{\text{counterfactual}} \tag{1}$$

where the key challenge is to construct the unobserved counterfactual and estimate $\theta$ accordingly. To reconstruct the missing counterfactual scenario, we rely on the latent factor model that accommodates pre-$T_0$ outcomes and auxiliary covariates to estimate the unobserved counterfactual component:

$$Y_{i,t}^N = \eta_t + \pi_t \cdot Z_i + \mu_t \cdot \phi_i + \varepsilon_{i,t} \tag{2}$$

---
[3] Iraq, Israel, Jordan, Lebanon, Turkey

where $\eta_t$ is an unobserved common factor that mimics time-fixed effects, $Z_i \in \mathbb{R}^r$ is a vector of observed time-varying and time-invariant covariates unaffected by the civil war, $\pi'_t \in \mathbb{R}^r$ is a vector of known parametric factor loadings, and $\mu'_t \in \mathbb{R}^F$ is a vector of common unobserved factors, and $\phi'_t \in \mathbb{R}^F$ is a vector of unknown factor loadings. The term $\varepsilon_{i,t}$ denotes transitory outcome shocks with $\varepsilon_{i,t} \sim \mathbb{N}(0,1)$ form of distribution. The key advantage of the latent factor model is to allow heterogeneous response of outcomes to multiple unobserved factors ($\mu_t \cdot \phi_i$) and embeds time trend models therein. The proposed latent factor model implicitly assumes that the number of common unobserved factors ($\mu_t$) is fixed over time which invokes the absence of structural breaks. The basic intuition behind the latent factor model is to reweigh the control group so that the synthetic version of Syria will match its observed $Z_i$ and some $Y_{i,t<T_0}$. Hence, $\phi_i$ will be matched by default and the set of unobserved common factors will not be projected out of the outcome generating process allowing for heterogeneous response to multiple unobserved factors. Consider a simple $J \times 1$ single dimensional vector of weights $W = (w_2, \dots, w_{J+1})'$ where $w_j \geq 0$ for $j = 2, \dots, J+1$ and $\sum_{j=2}^{J+1} w_j = 1$. Notice that each particular value of $W$ represents the weighted average of the implicit outcome- and covariate-related characteristics of the countries from a donor pool where $w_j \neq 0$. Let $\mathbf{X}$ represent pre-$T_0$ outcomes and covariates combined into a single vector that is partitioned into $\mathbf{X}_{Syria}$ and $\mathbf{X}_0$ where the former captures the values of matching variables for Syria and $\mathbf{X}_0$ represents the values of the variables for the countries unaffected by the civil war. To appropriately measure the discrepancy between Syria and the unaffected countries, we estimate $W$ through a nested optimization method using constrained quadratic programming routine to find best-witting set of weights conditional on the matrix of covariates. It is based on the algorithm using Vanderbei (1999) interior point method to solve the constrained quadratic programming problem, and is implemented via C++ plugin with 5 percent margin for the constraint violation tolerance using high-speed maximum likelihood approach as our default. We use a standard inner optimization to minimize a simple Euclidean distance between $\mathbf{X}_{Syria}$ and $\mathbf{X}_0$ to find the best-fitting weight set:

$$W^* = \underset{W}{\operatorname{argmin}} \|\mathbf{X}_{Syria} - \mathbf{X}_0 W\|_V = \sqrt{(\mathbf{X}_{Syria} - \mathbf{X}_0 W)' \mathbf{V} (\mathbf{X}_{Syria} - \mathbf{X}_0 W)} \qquad (3)$$

where $V$ is a symmetric diagonal matrix with positive components wherein $k$ diagonal elements $(v_1, \ldots v_k)$ represent the predictive weights of the pre-$T_0$ fitted variables. In the next step, in the outer optimization, $V$ can be estimated to minimize mean square prediction error in the pre-war period such that:

$$V^* = \underset{V}{\mathrm{argmin}} \left(\mathbf{Y}_{Syria} - \mathbf{Y}_0 \mathbf{W}^*(\mathbf{V})\right)' \left(\mathbf{Y}_{Syria} - \mathbf{Y}_0 \mathbf{W}^*(\mathbf{V})\right) \tag{4}$$

where $\mathbf{Y}_{Syria}$ denotes pre-war outcomes of Syria and $\mathbf{Y}_0$ captures a variety of linear combinations of pre-civil war outcomes of the unaffected countries that represent potential synthetic control units. To ensure that optimal non-negative weights exist as a convex linear combination of control countries, it is necessary that the interpolation biases are not present. As noted by Bluzscs and Valente (2022), such biases may occur if the synthetic control estimator obtains weights for countries that differ substantially from the war-affected countries in terms of unobservable time-varying confounding influences that shape both the outcome itself and the probability of armed conflict. We partially mitigated the interpolation biases by excluding all countries that have experienced armed internal conflicts during the period of our investigation. In addition, the exclusion of neighbouring countries that have been affected by Syrian civil war vis-á-vis the spillover effects such as increased refugee inflows and reduced economic activity, or have somewhat anticipated the war further bolsters the satisfaction of stable unit treatment value assumption, ensuring that neither the anticipation nor the spillover effects are present in the latent factor model that sets out to reconstruct $Y_{i,t}^N$. Under the regularity of these conditions, the average treatment effect of civil war on Syria can be written as:

$$\hat{\theta}_1 = \frac{1}{(T-T_0)} \cdot \sum_{t>T_0} \left(Y_{1t} - \sum_{j=2}^{J+1} w_j^* Y_{j,t}^N\right) \tag{5}$$

Furthermore, the synthetic representation of Syria's outcome in the pre-war period reinforce the estimated unobserved counterfactual from the latent factor provided that the set of weights is identified in the inner and outer optimization which implies that the outcome based on reweighting the synthetic control group is as follows:

$$Y_{W,t} = \sum_{j=2}^{J+1} w_j \cdot Y_{j,t} = \eta_t + \pi_t \cdot \left(\sum_{j=2}^{J+1} w_j Z_j\right) + \mu_t \cdot \left(\sum_{j=2}^{J+1} w_j \phi_j\right) + \left(\sum_{j=2}^{J+1} w_j \varepsilon_{j,t}\right) \tag{6}$$

provided that pre-war period is sufficiently large and that interpolation bias issues are properly mitigated, the synthetic control estimator provides a plausible representation of the counterfactual scenario in response to the civil war under time-varying heterogeneity. It should be noted that we divide the pre-treatment period into the training and validation period to choose ***V*** such that the resulting synthetic control region can plausibly approximate the outcome trajectory before the civil war. Provided that the number of pre-war period periods is large enough, for a given ***V***, ***W*** can be computed directly using the covariate matrix from the training period to minimize the mean squared prediction error during the validation period when the set of weights is produced.

*III.B Inference*

Since large-sample asymptotic inference is not possible with synthetic control method, Abadie et. al. (2010) advocate the use of treatment permutation method for inference on the treatment effect of interest. In our setup, we undertake a simple treatment permutation in space to detect whether the effect of the civil war on the growth and development outcomes is statistically significant at conventional levels. Treatment permutation consists of the iterative assignment of the civil war shock to the unaffected countries and use the outcome gaps to build the appropriate distribution of placebo effect. We proceed in two steps. In the first step, we estimate the placebo treatment effect by assigning the 2011 civil war shock to each unaffected country in the donor pool. In the second step, we compute the fraction of countries having the placebo effect at least as large as that of Syria. Our intuition is simple and straightforward. If the effect of civil war is both negative and statistically significant, the negative outcome gap should be unique to Syria and not perceivable elsewhere in the donor pool. By contrast, if the effect of civil war is not statistically significant, the placebo distribution should indicate no difference in the estimated effect of war for Syria and the placebo runs. Therefore, in step two, we calculate the empirical p-value for effect of civil war on Syria from a simple two-tailed empirical distribution:

$$\mathbb{P}_{i,t} = \frac{\sum_{j=2}^{J+1} \mathbf{1} \cdot \{\hat{\theta}_{j \in J+1} \geq \hat{\theta}_1\}}{J}$$

where $\mathbb{p}_{i,t}$ denotes two-tailed empirical p-value for the estimated dynamic treatment effect of civil war, $\hat{\theta}_{j\in J+1}$ denotes the full-sample placebo effect, $\hat{\theta}_1$ designates the estimated treatment effect of war on Syria, and $J$ indicates the size of the donor pool. Notice that the empirical p-value represents the probability to obtain dynamic average treatment effect at least as large as the one for Syria. However, if the prediction accuracy in the placebo simulation is low, the null hypothesis on the treatment effect of civil war is prone to over-rejection given a relatively large rarity of obtaining a large placebo effect. To address the type-II error, we compute pre-war average prediction error and constrain the inference procedure to ensure that p-values are estimated only if the prediction error in each placebo simulation is lower or equal to the mean prediction error for Syria (Seifert and Valente 2017, Ferman and Pinto 2017, Firpo and Possebom 2018, Arkhangelsky et. al. 2019) Furthermore, we discard the countries with mean prediction error at least four times that of Syria to partially eliminate the possibility of under-rejecting the null hypothesis that may arise if poorly fit placebos were included in the simulation. Our intuition behind the intertemporal behaviour of empirical p-values is two-fold. First, if the accuracy-adjusted p-values are consistently low from the early years of conflict onward, the analysis may highlight the civil war as a source of persistent breakdown in growth and development outcomes. And second, if the adjusted p-values by the end of the sample are high, then the notion of civil war as a temporary negative shock with few long-term implications becomes more obvious. Albeit imperfect by default, such inference procedure allows us detect whether the civil war in Syria has led to a permanent derailment of economic growth and development trajectories to evaluate the significance of the effect in greater detail.

## IV Data

### IV.A Dependent variables

Our set of dependent variables to capture economic growth, human development and institutional quality consists of a series of outcomes. First, we proxy the trajectory of economic growth using the GDP per capita denoted in Geary-Khamis international dollar at 2017 prices using recently updated series from Penn World Tables 10.0 (Feenstra et. al. 2015). Second, our proxy for the degree of social and economic development is the human development index and comes from United Nations Human Development Reports. More specifically, the index is a static composite index of life expectancy, mean years of schooling and per capita income. A higher score indicates longer lifespan, more affluent levels of education and higher per capita income whilst the index is in the range between zero and one. In addition, we consider infant mortality and life expectancy at birth as separate outcomes in the analysis to disentangle the effect of civil war on specific dimensions of human development. The data on life expectancy at birth is from World Health Organization. The rate of infant mortality is measured the number of child deaths under 5 years of age per one thousand births and comes from the updated *World Development Indicators*. Third, to capture the effect of war on human capital, we proxy human capital investment using the composite index of human capital that combines the mean years of education (Barro and Lee 2013) and the returns to human capital (Cohen and Soto 2007, Cohen and Leker 2014). And fourth, to capture the effect of war on the costs and extent of social cooperation, we use the population density rate as a rough proxy thereof. The density is denoted in the number of resident per squared km2 of land area, and provides a plausible insight into the extent of displacement induced by the civil war.

Another layer of dependent variables comprises the institutional quality outcomes. To estimate the effect of civil war on the trajectory of institutional quality over time, we use the indicators of quality from updated *Worldwide Governance Indicators* (Kaufmann et. al. 2008). First, voice and accountability index reflects the perception of the degree to which the citizens of the country are able to participate in free and fair regular elections as well as the degree of civil liberties. Second, political stability and absence of violence reflects the likelihood that the government will be destabilized or overthrown by violent and unconstitutional means that include politically-motivated violence and terrorism. Third, government effectiveness index measure the quality of public services, civil service as well as the quality of policy formulation, selection, implementation and the credibility of governments, and its commitment to raise and uphold high-quality policy-making. Fourth, somewhat relatedly, regulatory quality index captures the perceptions of the ability of the government to formulate and implement sound economic policies and regulations that facilitate private-sector development. Fifth, rule of law index reflects the perceptions of the degree to which citizens have confidence in and abide by the rule of society as well as the quality of contract enforcement, security of property rights for non-elites, the reliability of police and quality of the judicial system in resolving disputes as well as the likelihood of crime and violence. And sixth, we also use the control of corruption index. This particular indicator measures the extent to which public power is abused for private gain, including both petty and grand forms of corruption as well as the capture of the state by the political and business elites and the private interest groups.

*IV.B Auxiliary covariates*

The auxiliary covariates used to match Syria's salient growth and development characteristics with the rest of the world are grouped into several blocks. The first block comprises pre-war outcomes in key benchmark years to capture the potential state dependence in each outcome trajectory. The second block comprises physical geographic matching variables which are independent of the dynamic outcome realization. These variables include terrain ruggedness, latitude, longitude, soil quality, fraction of desert area, fraction of tropical area, distance from coastline, accessibility of coastline and continental indicator of Asia where Syria is located. The variables come from Nunn and Puga (2012). The third block comprises a dichotomous variable, indicating whether the countries belong to the civil law tradition (La Porta et. al. 1998).

*IV.C Treatment and control samples*

Our treatment sample consists of Syria and our time period of investigation is between the years 1996 and 2021. Drawing on the updated dataset of armed conflict (Brecke 1999), we build a donor pool of 66 countries[4] where no armed internal conflict took place in the period 1996-2021. Our treatment and control samples are strongly balanced. It should be noted that neighbouring countries receiving spillover effects of Syria's civil war through the inflow of refugees and the associated humanitarian costs are discarded from the donor pool. Due to the likely violated of stable unit treatment value assumption (SUTVA), it is highly likely that the inclusion

---

[4] Albania, Argentina, Armenia, Australia, Austria, Azerbaijan, Bahrain, Belarus, Belgium, Bolivia, Bosnia and Herzegovina, Botswana, Brazil, Bulgaria, Cape Verde, Cambodia, Canada, Chile, China, Costa Rica, Croatia, Cyprus, Czech Republic, Denmark, Dominican Republic, Ecuador, Estonia, Finland, France, Germany, Greece, Honduras, Hungary, Iceland, Ireland, Italy, Japan, Latvia, Lithuania, Luxembourg, Malaysia, Malta, Namibia, Netherlands, New Zealand, Nicaragua, North Macedonia, Oman, Poland, Portugal, Qatar, Republic of Korea, Russia, Serbia, Singapore, Slovakia, Slovenia, South Africa, Spain, Sweden, Switzerland, United Arab Emirates, United Kingdom, United States, Uruguay, Venezuela

of the neighbouring countries into the donor pool would conflate the treatment effect of civil war with severe biases and possibly lead to implausible estimate of the dynamic treatment effect.

## V    Results

*V.A Effect of civil war on economic growth and social development*

Figure 1 reports the estimated effect of civil war on the trajectory of Syria's economic growth and social development. To capture economic growth and some of its mechanisms, several variables are used in the empirical analysis through the application of synthetic control estimator. The evidence invariably suggests large losses of per capita GDP along with a rampant deterioration of social development in the years after the civil war. Without the loss of generality, the key insights can be drawn from the analysis of the effects of war on Syrian economy. First, in the light of the general effects of civil war reported in the existing literature (Collier and Sambanis 2005), Syrian civil war appears to be a temporary shock for the trajectory of economic growth with no discernible long-term effect on per capita GDP. That said, the trajectory of growth uncovers a steady and rapid deterioration of output in the first three years of the war. After 2013, Syria's GDP appears to rebound and gradually approach the trajectory of its synthetic control group before the onset of COVID19 pandemic. In the end-of-sample year, our estimates suggest that Syria's GDP per capita is 813 USD below the bar of its synthetic control group, or roughly 12 percent relative to the level of its per capita GDP. By iteratively permuting the civil war to the unaffected countries through in-space placebo analysis (Galiani and Quistorff 2018), the p-value on the effect of civil war by the end of our sample period is roughly 0.520 which implies that the war inflicted a deep but temporary shock onto the trajectory of growth with no evidence of permanent structural breakdown compared to the effect of war in former Yugoslavia (Keseljevic and Spruk 2022). The economic growth trajectory of Syria before the civil war is best reproduced as a convex combination of countries' growth and development attributes that fall within its convex hull such as Honduras (73 percent), Bolivia (18 percent), Belgium (5 percent), and Cambodia (4 percent), respectively.

Second, the effect of the civil war on social development appears to be particularly devastating and mimics the characteristics of the structural breakdown. For instance, our estimates suggest that the civil war has stymied the human development index by 0.15 basis points compared to its synthetic control group. The end-of-sample classical p-values on the effect of civil war with respect to human development is 0.000 and confirms statistically significant and very large negative effect. Compared to its synthetic control group, the trajectory of human development exhibits a large downward sloping trend with sign of the mild reversal in 2017 followed by a stagnation in the year preceding the COVID19 pandemic. Our synthetic control estimates highlight a pattern of steady growth and reinforced improvement in human development index in the hypothetical absence of the war. Pre-war trajectory of Syria's human development index can be best reproduced as a convex combination of the implied attributes of Serbia (33 percent), Cambodia (33 percent), Armenia (30 percent), South Africa (4 percent), and Bulgaria (<1 percent). A pervasive drop in human development appears to be permanent with no sign of reversal that would indicate a long-lasting but temporary shock to human development induced by the civil war. By examining the effect of civil war on specific components of human development, it becomes apparent that the negative effect of civil war on human development materializes mainly through a drop of longevity rather than human capital deterioration. For instance, before the civil war, Syria's life expectancy at birth was around 74 years. Until 2015, life expectancy decreased to 69 years which entails around one life year deterioration for each year of the armed conflict. Our synthetic control estimates invariably suggest that in the hypothetical absence of the war, life expectancy trajectory would reach 78 years. In comparative perspective, such life expectancy threshold is comparable with Estonia and United States. By 2019, our estimates thus imply that Syria's life expectancy is around 5.12 years lower compared to its synthetic control group. The end-of-sample p-value of the effect of civil war is 0.000 and readily demonstrate a large-scale significance of civil war in the deterioration of longevity during the armed conflict. The composition of synthetic control group invariably suggests that pre-war trajectory of Syria's life expectancy is best reproduced by the implicit attributes of Serbia (32 percent), Netherlands (24 percent), Oman (22 percent), Uruguay (16 percent), Cambodia (5 percent) and Bolivia (2 percent), respectively. On the contrary, the effect of civil war on human capital appears to be weak. The trajectory of human capital index of Syria is almost identical to the trajectory of its synthetic control group before and after the war with minimal predictive discrepancy (i.e. RMSE = 0.005). The null hypothesis on the human capital effect of war cannot be rejected at conventional levels (i.e. p-value = 0.391), further reinforcing the notion of almost zero effect of war on human capital. By contrast, the effect of civil war on infant mortality appears to be very large. Our synthetic control estimates invariably suggest that in the hypothetical absence of the war, infant mortality would continually exhibit a downward trajectory reaching the level of 1.5 deaths per 1,000 live births which is comparable with Ecuador by the end of our sample period. By 2019, the observed rate of infant mortality reached 2.15 deaths per 1,000 live births which is one of the highest rates of mortality in our sample and is comparable with Venezuela. The end-of-sample classical permutation-based p-value on the infant mortality effect of civil war is 0.000 which confirms a permanent increase in mortality toll after the civil war. The trajectory of Syria's infant mortality prior to the civil war is best reproduced as a convex combination of the implicit characteristics of Malta (36 percent), Dominican Republic (36 percent), Oman (21 percent), Cape Verde (5 percent), Bulgaria (2 percent) and Serbia (<1 percent), respectively. In real terms, our quantitative mortality estimates imply that the infant mortality toll of the Syrian civil war amounts to 5,700 additional infant deaths relative to its synthetic control group where no armed conflict took place and resembles similar implicit trajectory of infant mortality to Syria.

**Figure 1**: Economic growth and social development effect of civil war on Syrian economy, 1995-2021

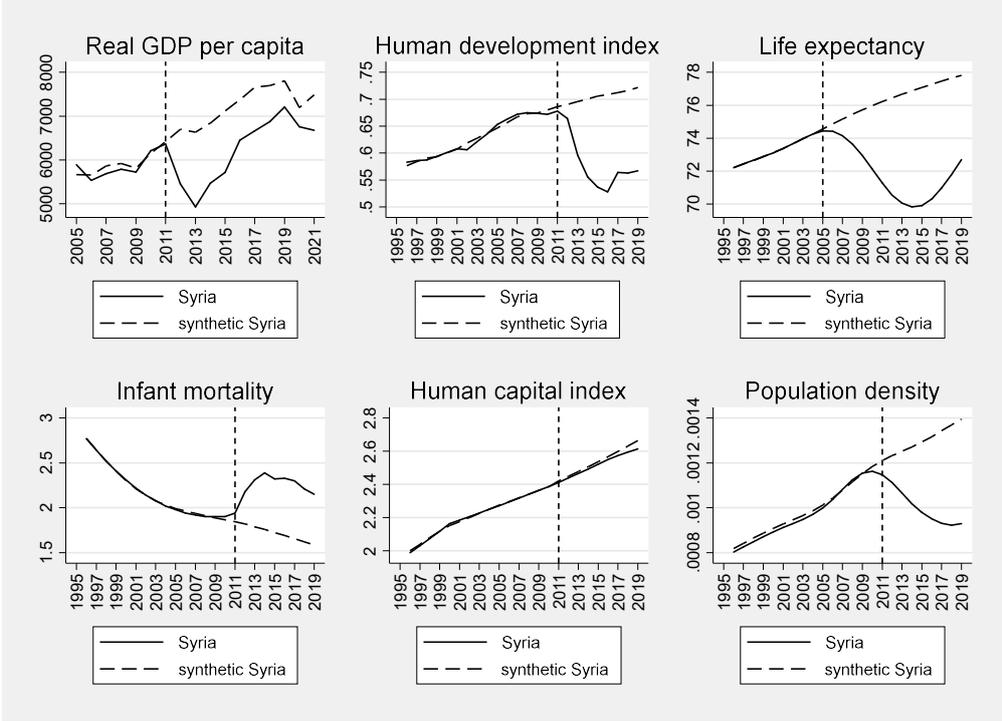

In addition, our analysis uncovers evidence of large-scale population displacement. Although a more nuanced analysis would entail the use of data on the outflow of migrants, our approach is to capture the effect of war on displacement using population density rate and estimate the appropriate counterfactual scenario. The evidence suggests that a synthetic Syria provides an excellent fit of the population density trajectory in the period before the civil war. The discrepancy between the observed density trajectory and its synthetic peer appears to be minimal (i.e. RMSE = 0.0001). Our synthetic control estimate indicate a continuously increasing trajectory of population density in the hypothetical absence of the war. After the onset of the civil war, population density exhibits a persistent decrease that appears to be permanent. By the end of our sample, the estimated population density gap is around 2.71 inhabitants per 1000 km2 which entails the reduced of almost three individuals per 1000 km2 or the annual loss of slightly more than 500,000 people for each year of the duration of war. The population density trajectory of Syria before the civil war is best reproduced as a convex combination of demographic and auxiliary characteristics of Cambodia (44 percent), Honduras (19 percent), Argentina (18 percent), Oman (11 percent), Cape Verde (4 percent), Russia (3 percent), Bahrain (1 percent) and Singapore (1 percent), respectively. Table 1 reports baseline covariate balance along with a more detailed composition of synthetic control groups and exposition of the structural effect of civil wars for each outcome variable in our investigation.

Table 1: Covariate Balancing and Composition of Synthetic Control Groups

| | Real GDP per capita | Human development index | Life expectancy | Infant mortality | Human capital formation | Population density |
|---|---|---|---|---|---|---|
| Pre-treatment training and validation period | 2005-2010 | 1995-2010 | 1995-2005 | 1995-2010 | 1995-2010 | 1995-2010 |
| Pos-treatment period | 2011-2021 | 2011-2021 | 2005-2021 | 2010-2021 | 2010-2021 | 2010-2021 |
| End-of-sample delta | -813 USD | -0.154 basis points | -5.12 years | +0.57 deaths per 10,000 births | -0.064 basis points | -2.71 inhabitants per km2 |
| RMSE | 141.32 | 0.005 | 0.010 | 0.013 | 0.005 | 0.0001 |
| End-of-sample classical p-value | (0.520) | (0.000) | (0.000) | (0.000) | (0.391) | (0.000) |
| Albania | 0 | 0 | 0 | 0 | 0 | 0 |
| Argentina | 0 | 0 | 0 | 0 | 0 | 0.18 |
| Armenia | 0 | 0.30 | 0 | 0 | 0 | 0 |
| Australia | 0 | 0 | 0 | 0 | 0 | 0 |
| Austria | 0 | 0 | 0 | 0 | 0 | 0 |
| Azerbaijan | 0 | 0 | 0 | 0 | 0 | 0 |

| Country | | | | | | |
|---|---|---|---|---|---|---|
| Bahrain | 0 | 0 | 0 | 0 | 0 | 0.01 |
| Belarus | 0 | 0 | 0 | 0 | 0 | 0 |
| Belgium | 0.05 | 0 | 0 | 0 | 0 | 0 |
| Bolivia | 0.18 | 0 | 0.02 | 0 | 0 | 0 |
| Bosnia and Herzegovina | 0 | 0 | 0 | 0 | 0 | 0 |
| Botswana | 0 | 0 | 0 | 0 | 0 | 0 |
| Brazil | 0 | 0 | 0 | 0 | 0.33 | 0 |
| Bulgaria | 0 | <0.01 | 0 | 0.02 | 0 | 0 |
| Cape Verde | 0 | 0 | 0 | 0.05 | 0 | 0.04 |
| Cambodia | 0.04 | 0.33 | 0.05 | 0 | 0.25 | 0.44 |
| Canada | 0 | 0 | 0 | 0 | 0 | 0 |
| Chile | 0 | 0 | 0 | 0 | 0 | 0 |
| China | 0 | 0 | 0 | 0 | 0 | 0 |
| Costa Rica | 0 | 0 | 0 | 0 | 0 | 0 |
| Croatia | 0 | 0 | 0 | 0 | 0 | 0 |
| Cyprus | 0 | 0 | 0 | 0 | 0 | 0 |
| Czech Republic | 0 | 0 | 0 | 0 | 0 | 0 |
| Denmark | 0 | 0 | 0 | 0 | 0 | 0 |
| Dominican Republic | 0 | 0 | 0 | 0.36 | 0 | 0 |
| Ecuador | 0 | 0 | 0 | 0 | 0 | 0 |
| Estonia | 0 | 0 | 0 | 0 | 0 | 0 |
| Finland | 0 | 0 | 0 | 0 | 0 | 0 |
| France | 0 | 0 | 0 | 0 | 0 | 0 |
| Germany | 0 | 0 | 0 | 0 | 0 | 0 |
| Greece | 0 | 0 | 0 | 0 | 0 | 0 |
| Honduras | 0.73 | 0 | 0 | 0 | 0 | 0.19 |
| Hungary | 0 | 0 | 0 | 0 | 0 | 0 |
| Iceland | 0 | 0 | 0 | 0 | 0 | 0 |
| Ireland | 0 | 0 | 0 | 0 | 0 | 0 |
| Italy | 0 | 0 | 0 | 0 | 0 | 0 |
| Japan | 0 | 0 | 0 | 0 | 0 | 0 |
| Latvia | 0 | 0 | 0 | 0 | 0 | 0 |
| Lithuania | 0 | 0 | 0 | 0 | 0 | 0 |
| Luxembourg | 0 | 0 | 0 | 0 | 0 | 0 |
| Malaysia | 0 | 0 | 0 | 0 | 0 | 0 |
| Malta | 0 | 0 | 0 | 0.36 | 0 | 0 |
| Namibia | 0 | 0 | 0 | 0 | 0 | 0 |
| Netherlands | 0 | 0 | 0.24 | 0 | 0 | 0 |
| New Zealand | 0 | 0 | 0 | 0 | 0 | 0 |
| Nicaragua | 0 | 0 | 0 | 0 | 0 | 0 |
| North Macedonia | 0 | 0 | 0 | 0 | 0 | 0 |
| Oman | 0 | 0 | 0.22 | 0.21 | 0 | 0.11 |
| Poland | 0 | 0 | 0 | 0 | 0 | 0 |
| Portugal | 0 | 0 | 0 | 0 | 0 | 0 |
| Qatar | 0 | 0 | 0 | 0 | 0 | 0 |
| Republic of Korea | 0 | 0 | 0 | 0 | 0 | 0 |
| Russia | 0 | 0 | 0 | 0 | 0 | 0.03 |
| Serbia | 0 | 0.33 | 0.32 | <0.01 | 0 | 0 |
| Singapore | 0 | 0 | 0 | 0 | 0 | 0.01 |
| Slovakia | 0 | 0 | 0 | 0 | 0 | 0 |
| Slovenia | 0 | 0 | 0 | 0 | 0 | 0 |
| South Africa | 0 | 0.04 | 0 | 0 | 0 | 0 |
| Spain | 0 | 0 | 0 | 0 | 0 | 0 |
| Sweden | 0 | 0 | 0 | 0 | 0 | 0 |
| Switzerland | 0 | 0 | 0 | 0 | 0 | 0 |
| United Arab Emirates | 0 | 0 | 0 | 0 | 0.41 | 0 |
| United Kingdom | 0 | 0 | 0 | 0 | 0 | 0 |
| United States | 0 | 0 | 0 | 0 | 0 | 0 |
| Uruguay | 0 | 0 | 0.16 | 0 | 0 | 0 |
| Venezuela | 0 | 0 | 0 | 0 | 0 | 0 |

*V.B Effect of civil war on institutional quality*

Figure 2 reports the effect of civil war on the trajectory of institutional quality. The main dependent variable to capture the institutional quality is the first principal of component of updated Kaufmann et. al. (2011) governance indicators[5] that partially reflect the quality and strength of public governance and institutional framework. The evidence from our synthetic control analysis unveils persistent and rampant deterioration of institutional quality after the war towards more extractive and exclusionary institutional framework titled in favor of the ruling elite network. In search of the missing counterfactual scenario, synthetic control estimator provides a reasonably good quality of the institutional quality fit between Syria and its synthetic control group. The estimated institutional quality gap after the beginning the civil war is both large and negative, indicating a pervasive worsening and weakening of the institutional framework in both enshrining and ensure the rule of law, effective functioning of government administration and safeguarding the control over corruption whilst failing to uphold the strength and freedom of the civil society in monitoring the elites in power. Our evidence arguably shows no tendency of institutional quality recovery in the years after the beginning of the civil war. In particular, end-of-sample gap estimates show that the Syria's the first principal component of institutional quality dropped by 2.8 basis points in comparison with its synthetic control group. The negative effect of the civil war on the overall institutional quality conditions appears to be strong and statistically significant (i.e. p-value = 0.000). The negative effect appears to unfold gradually and tends to amplify by the end of our sample period. Furthermore, Syria's trajectory of institutional quality's first principal component is best reproduced by the convex combination of the quality attributes of Russia (29 percent), Azerbaijan (28 percent), Venezuela (23 percent) and Cambodia (20 percent) which fall within the convex hull of Syria's institutional quality path in the pre-war training and validation period.

---

[5] The baseline indicators used to construct the first principal component of institutional quality include (i) voice and accountability, (ii) political stability and absence of violence, (iii) regulatory quality, (iv) government effectiveness, (v) rule of law and (vi) control of corruption.

Figure 2: Effect of civil war on institutional quality of Syria, 1995-2020

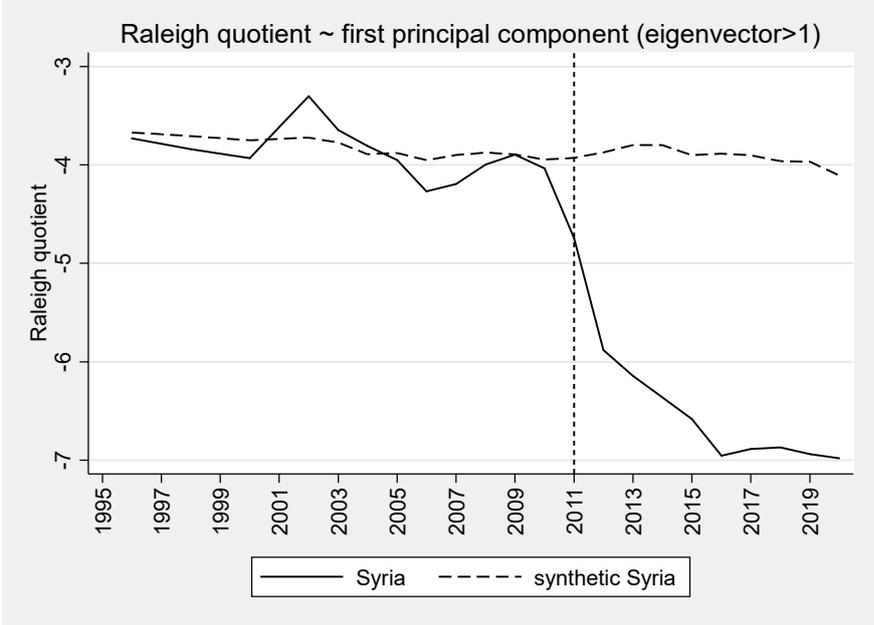

In terms of further example, we examine the effect of the civil war beyond the general realm of institutional quality and estimate the gaps with respect to each quality variable separately. Table 2 reports the effect of civil war on six separate dimensions of institutional quality and the first principal component of the respective dimensions. The estimated gaps invariably uncover evidence of pervasive and prolonged deterioration of institutional quality along each dimension in varying degrees. The effect of civil war on the ability of citizens to participate in government selection, freedom of expression and association has been substantial although not overly large in the light of Syria's low score on the freedom of expression metrics prior to the war. Synthetic control estimates indicate a prolonged deterioration in voice and accountability index after the war although the effect does not appear to be statistically significant (i.e. end-of-sample p-value = 0.227). By contrast, the effect of war on political stability and state-mandated violence has been particularly devastating. Our end-of-sample point estimate uncovers a large political stability and violence gap in response to the civil war. In quantitative terms, our estimates show that the political stability and violence index is derailed by 2.2 basis points by the end of the sample period with a clear indication of statistically significant effect at conventional level (i.e. p-value = 0.000). Syria's trajectory of political stability and violence before the civil war can be best synthesized as convex combination of the stability and peace attributes of Bahrain (36 percent), Greece (25 percent), Russia (24 percent), Venezuela (9 percent), Bolivia (5 percent) and Bosnia and Herzegovina (1 percent), respectively.

The general thrust of our estimates emphasizes a pervasive deterioration of institutional quality and governance conditions. Furthermore, the effect of the civil war is arguably devastating in terms of considerably less effective government administration and the quality of regulation. Our point estimates convey a clear signal of the pervasive drop in the quality of public services, civil service, policy formulation and policy implementation amidst a shackled credibility of government's commitment to maintain these qualities. Our estimates of the treatment effect of civil war on government effectiveness indicate 1.15 basis point drop in the respective index by 2020 which is arguably large and statistically significant (i.e. p-value = 0.000). In terms of the composition of synthetic control group, the effectiveness attributes of countries such as Cambodia, Azerbaijan, Serbia and Ecuador are best able to track and reproduce the government effectiveness trajectory of Syria prior to the onset of the civil war. Similar negative effects are apparent with respect to the quality of regulation where our end-of-sample point estimates highlight 0.99 basis point drop that appears to be permanent (i.e. p-value = 0.015). The regulatory quality trajectory of Syria is best tracked by a convex combination of the implied characteristics of Belarus (53 percent), Azerbaijan (27 percent), and Venezuela (20 percent), respectively. Furthermore, the effect of the civil war on the rule of law is particularly large and calamitous. A drop in the rule of law after the civil war is particularly large and around 1.65 basis points by the end of the sample period in 2020 and also appears to be permanent (i.e. p-value = 0.000). A drop in the rule of law indicates a decrease in the quality of contract enforcement, weakening of the security of property rights, lower trust in the police and undermined credibility of the courts in addition in strongly increased likelihood of crime and violence after the beginning of the civil war. The set of countries in the synthetic control group that best track and reproduce Syria's rule of law and fall within its convex hull comprises Armenia (65 percent), Bolivia (17 percent) and Cambodia (17 percent), respectively.

Aligned with our theoretical expectation, we also find evidence of heightened prevalence of corruption in the years after the civil war. The effect of the civil war on the prevalence of corruption is moderately large and indicates a significantly increased exercise of public power by the existing political regime for private gain along with the capture of the state by the elites and their private interests. By permuting the civil war to the unaffected countries, the effect of civil war on the control of corruption is marginally significant (i.e. p-value = 0.091) and somewhat lower in comparison with the devastating effect of war on political stability, rule of law and government effectiveness. The trajectory of control of corruption index of Syria prior to the civil war is best synthesized as a convex combination of the implicit characteristics of Cambodia (47 percent), Nicaragua (26 percent), Azerbaijan (14 percent), Armenia (7 percent), and Russia (6 percent). Table 2 reports more detailed results along with the composition of synthetic control group. Figure 3 presents the overall frequency of non-zero weight in the donor pool for growth and development outcomes and institutional quality outcomes and parses out the most influential contributing countries in the variety of synthetic control groups for Syria that are able to approximate a plausible trajectory of economic and institutional development in the hypothetical absence of the war. We also perform a series of robustness check such as leave-one-out analysis, in-time placebo analysis and the analysis with a more restrictive donor pool that compares Syria to the countries with salient institutional features, and show that our baseline effects of civil war on economic growth, human development and institutional quality are very robust to these checks.[6]

Table 2: Effect of civil war on the institutional development of Syria and the composition of synthetic control groups, 1996-2020

|  | Voice and accountability | Political stability and absence of violence | Government effectiveness | Regulatory quality | Rule of law | Control of corruption | Raleigh Quotient – first principal component |
|---|---|---|---|---|---|---|---|
| Pre-treatment training and validation period | 1995-2010 | 1995-2010 | 1995-2005 | 1995-2010 | 1995-2010 | 1995-2010 | 1995-2010 |
| Post-treatment period | 2011-2020 | 2011-2020 | 2011-2020 | 2011-2020 | 2011-2020 | 2011-2020 | 2011-2020 |

---

[6] For the sake of space limitation, these robustness checks are not reported. However, we are very happy to provide for evaluation and review upon the request of referees and/or the editors.

| | | | | | | | |
|---|---|---|---|---|---|---|---|
| End-of-sample delta | -0.286 basis points | -2.211 basis points | -1.147 basis points | -0.994 basis points | -1.655 basis points | -0.611 basis points | -2.858 basis points |
| RMSE | **0.100** | **0.129** | **0.159** | **0.092** | **0.095** | **0.159** | **0.193** |
| End-of-sample classical p-value | (0.227) | (0.000) | (0.000) | (0.015) | (0.000) | (0.091) | (0.000) |
| Albania | 0 | 0 | 0 | 0 | 0 | 0 | 0 |
| Argentina | 0 | 0 | 0 | 0 | 0 | 0 | 0 |
| Armenia | 0 | 0 | 0 | 0 | 0.65 | 0.07 | 0 |
| Australia | 0 | 0 | 0 | 0 | 0 | 0 | 0 |
| Austria | 0 | 0 | 0 | 0 | 0 | 0 | 0 |
| Azerbaijan | 0 | 0 | 0.30 | 0.27 | 0 | 0.14 | 0.28 |
| Bahrain | 0 | 0.36 | 0 | 0 | 0 | 0 | 0 |
| Belarus | 0 | 0 | 0 | 0.53 | 0 | 0 | 0 |
| Belgium | 0 | 0 | 0 | 0 | 0 | 0 | 0 |
| Bolivia | 0 | 0.05 | 0 | 0 | 0.17 | 0 | 0 |
| Bosnia and Herzegovina | 0 | 0.01 | 0 | 0 | 0 | 0 | 0 |
| Botswana | 0 | 0 | 0 | 0 | 0 | 0 | 0 |
| Brazil | 0 | 0 | 0 | 0 | 0 | 0 | 0 |
| Bulgaria | 0 | 0 | 0 | 0 | 0 | 0 | 0 |
| Cape Verde | 0 | 0 | 0 | 0 | 0 | 0 | 0 |
| Cambodia | 0 | 0 | 0.57 | 0 | 0.17 | 0.47 | 0.20 |
| Canada | 0 | 0 | 0 | 0 | 0 | 0 | 0 |
| Chile | 0 | 0 | 0 | 0 | 0 | 0 | 0 |
| China | 1 | 0 | 0 | 0 | 0 | 0 | 0 |
| Costa Rica | 0 | 0 | 0 | 0 | 0 | 0 | 0 |
| Croatia | 0 | 0 | 0 | 0 | 0 | 0 | 0 |
| Cyprus | 0 | 0 | 0 | 0 | 0 | 0 | 0 |
| Czech Republic | 0 | 0 | 0 | 0 | 0 | 0 | 0 |
| Denmark | 0 | 0 | 0 | 0 | 0 | 0 | 0 |
| Dominican Republic | 0 | 0 | 0 | 0 | 0 | 0 | 0 |
| Ecuador | 0 | 0 | 0.06 | 0 | 0 | 0 | 0 |
| Estonia | 0 | 0 | 0 | 0 | 0 | 0 | 0 |
| Finland | 0 | 0 | 0 | 0 | 0 | 0 | 0 |
| France | 0 | 0 | 0 | 0 | 0 | 0 | 0 |
| Germany | 0 | 0 | 0 | 0 | 0 | 0 | 0 |
| Greece | 0 | 0.25 | 0 | 0 | 0 | 0 | 0 |
| Honduras | 0 | 0 | 0 | 0 | 0 | 0 | 0 |
| Hungary | 0 | 0 | 0 | 0 | 0 | 0 | 0 |
| Iceland | 0 | 0 | 0 | 0 | 0 | 0 | 0 |
| Ireland | 0 | 0 | 0 | 0 | 0 | 0 | 0 |
| Italy | 0 | 0 | 0 | 0 | 0 | 0 | 0 |
| Japan | 0 | 0 | 0 | 0 | 0 | 0 | 0 |
| Latvia | 0 | 0 | 0 | 0 | 0 | 0 | 0 |
| Lithuania | 0 | 0 | 0 | 0 | 0 | 0 | 0 |
| Luxembourg | 0 | 0 | 0 | 0 | 0 | 0 | 0 |
| Malaysia | 0 | 0 | 0 | 0 | 0 | 0 | 0 |
| Malta | 0 | 0 | 0 | 0 | 0 | 0 | 0 |
| Namibia | 0 | 0 | 0 | 0 | 0 | 0 | 0 |
| Netherlands | 0 | 0 | 0 | 0 | 0 | 0 | 0 |
| New Zealand | 0 | 0 | 0 | 0 | 0 | 0 | 0 |
| Nicaragua | 0 | 0 | 0 | 0 | 0 | 0.26 | 0 |
| North Macedonia | 0 | 0 | 0 | 0 | 0 | 0 | 0 |
| Oman | 0 | 0 | 0 | 0 | 0 | 0 | 0 |
| Poland | 0 | 0 | 0 | 0 | 0 | 0 | 0 |
| Portugal | 0 | 0 | 0 | 0 | 0 | 0 | 0 |
| Qatar | 0 | 0 | 0 | 0 | 0 | 0 | 0 |
| Republic of Korea | 0 | 0 | 0 | 0 | 0 | 0 | 0 |
| Russia | 0 | 0.24 | 0 | 0 | 0 | 0.06 | 0.29 |
| Serbia | 0 | 0 | 0.07 | 0 | 0 | 0 | 0 |
| Singapore | 0 | 0 | 0 | 0 | 0 | 0 | 0 |
| Slovakia | 0 | 0 | 0 | 0 | 0 | 0 | 0 |
| Slovenia | 0 | 0 | 0 | 0 | 0 | 0 | 0 |
| South Africa | 0 | 0 | 0 | 0 | 0 | 0 | 0 |
| Spain | 0 | 0 | 0 | 0 | 0 | 0 | 0 |
| Sweden | 0 | 0 | 0 | 0 | 0 | 0 | 0 |
| Switzerland | 0 | 0 | 0 | 0 | 0 | 0 | 0 |
| United Arab Emirates | 0 | 0 | 0 | 0 | 0 | 0 | 0 |
| United Kingdom | 0 | 0 | 0 | 0 | 0 | 0 | 0 |
| United States | 0 | 0 | 0 | 0 | 0 | 0 | 0 |
| Uruguay | 0 | 0 | 0 | 0 | 0 | 0 | 0 |
| Venezuela | 0 | 0.09 | 0 | 0.20 | 0 | 0 | 0.23 |

**Figure 3**: Frequency of non-zero donor weight across the specifications

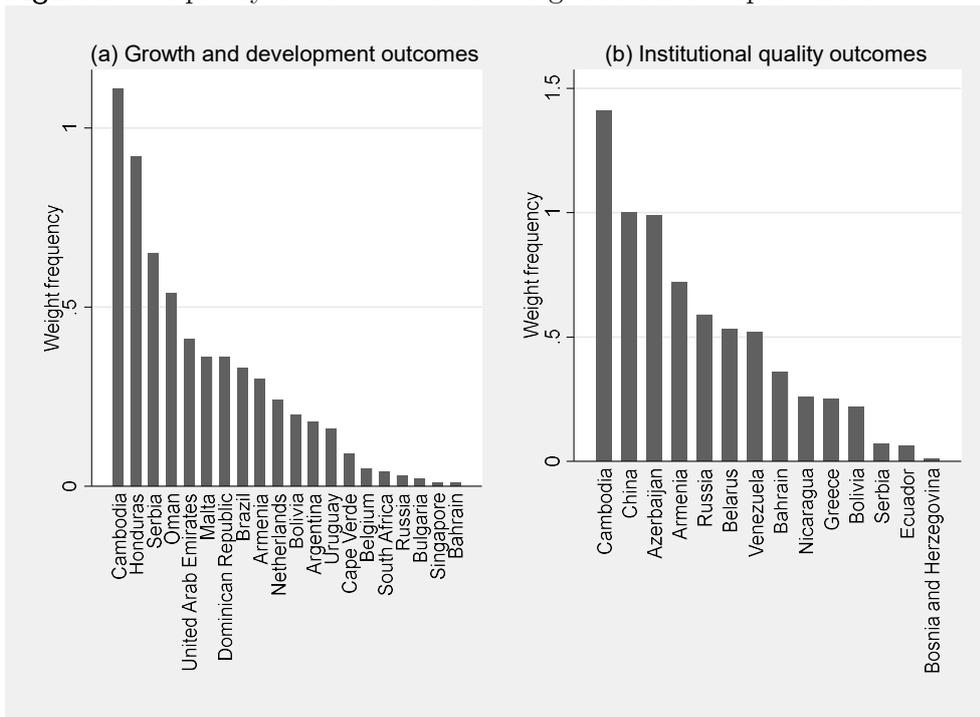

*V.3 Discussion*

Estimating the missing counterfactual growth and institutional development scenario parses out plausible trajectories of economic growth and institutional quality in the eventual absence of the civil war in Syria through the convex combination of growth and quality attributes of the countries directly unaffected by the civil war. Through the application of synthetic control estimator, the missing counterfactual scenario indicates pervasive and unprecedented losses of human development, particularly reduced longevity and substantially increased infant mortality. In addition, we find evidence of severe population displacement whilst the effect of war on per capita GDP appears to be temporary.

The obvious question behind the array of estimated gap concerns the comparative context of the results uncovered in our investigation in comparison with the previous findings in the literature on civil war. Similar to the armed conflicts in former Yugoslavia, sub-Saharan Africa, Colombia and Middle East (Collier and Sambanis 2005), the civil war in Syria appears to be a large-scale and negative shock with a sizeable and non-trivial effect on the trajectory of economic growth. Our synthetic control estimates show that the negative effect of war is particularly strong in the first two years of the conflict. After 2013, the growth trajectory of Syria unfolds into a recovery and down to the beginning of the COVID19 pandemic, the respective growth path almost fully converges with the synthetic control group that is dominated by countries such as Honduras and Bolivia. In contrast to the armed conflict in former Yugoslavia (Keseljevic and Spruk 2021), the negative effect of civil war on economic growth gradually weakens over time. Our point estimates that by 2021, Syrian per capita GDP would be 815 international dollars lower than it would be in the absence of the civil war. On a comparative ledger, such difference implies that had the civil war been avoided, Syria's per capita GDP down to the present day would be at the bar of Namibia and Bolivia. Given a large-scale drop in the investment rate and export share of GDP indicated by Penn World Tables (Feenstra et. al. 2015), it is plausible to note that the recovery has been driven by substantial increases in government consumption used to rebuild the damaged infrastructure, private domestic consumption and large-scale informal economy which has been estimated at around 20 percent of the GDP (Medina and Schneider 2018). Contrary to the popularized perception in the West, our results cast a shadow of doubt on the effectiveness of international sanctions imposed on Al Assad regime by European Union and United States that hinged on the prohibition of trade in goods and services, a series of embargos and freezing of financial assets among several others. Our evidence shows that, in terms of per capita GDP, the imposition of sanctions have not stifled the economic recovery after the formal ceasefire. On the other hand, the imposed sanctions on humanitarian and public health outcomes only seem to have prolonged the protracted and deep negative effects of war on various metrics of human development (Friberg Lyme 2012, Sen et. al. 2013, Moret 2015). One possible mechanism in counteracting deleterious effect of international sanctions may be the increased military and development assistance of countries such as China and, to a lesser extent, Russia and Iran (Burton et. al. 2021).

In terms of further example, the humanitarian cost of the civil war appears to be devastating. Our synthetic control estimates uncover evidence of considerably reduced longevity and massively increased infant mortality in the light of ailing and weakened public health capacity to provide the basic public goods and services to the civilian population. Up to the present day, the average life expectancy at birth in Syria is around 5 years lower than it would be in the hypothetical absence of the civil war. Through the lens of comparative perspective, our estimates imply that under no-war scenario, Syria's life expectancy in 2021 would be only one year lower compared to the United States. In terms of further example, before the civil war, Syria was project to reach the rate of infant mortality of the least developed EU member states in about 15 years under the pre-war demographic scenario. The civil war led to an unprecedented increase in infant mortality by 36 percent compared to the synthetic control group. In real terms, the quantitative estimates of infant mortality gap imply that the infant mortality toll amounts to 5,700 additional infant deaths for each year of the conflict compared to the control group of no armed conflict. Before the onset of the civil war and sectarian violence, Syria's human development index was comparable with China while we show that without the war, it would be 35 percent higher up to the present day whilst today it is comparable with countries such as Pakistan and Cameroon. Without the loss of generality, our estimates show that the civil war in Syria has destroyed and decimated an otherwise reasonably strong growth and development potential of a country, which would join the club of upper-middle income countries by 2025 in the absence of the armed conflict and the subsequent prolonged civil war.

Furthermore, our evidence also bolsters the notion that the civil war led to a more extractive and exclusionary institutional structure. In retrospect, comparing the evolution of the institutional quality trajectory with that of the estimated counterfactuals, it becomes apparent that the civil war led to a protracted suppression of civil liberties and freedom of expression, widespread state-sponsored violence against the civilian population and political opponents, weakened effectiveness of government administration, rampant deterioration of the rule of law, deleterious regulatory quality and more widespread corruption.

## VI    Conclusion

Civil unrest in Syria began in 2011 in a wave of Arab Spring protests in response to the discontent with Syrian government. Unlike elsewhere in the Middle East, the protests evolved and escalated into a massive armed conflict after the calls for the removal of Al Assad by Syrian opposition forces and the protestors were violently squashed. The war in Syria has been fought by several factions and quickly reached an international dimension The Syrian Armed Forces under the government control have been supported by Iran, Russia and Hezbollah while the opposition-controlled Syrian Democratic Forces and Syrian National Army received the support from United States and Turkey. Without the loss of generality, Syrian civil war has been characterized as one of the most violent episodes of armed conflict in modern history leading to the most widespread displacement of population and unprecedented refugee crisis and paramount humanitarian costs. In tackling the costs of civil war, the question that has received much less attention concerns the counterfactual dynamics, namely, how would Syrian economy, institutions and society have developed in the hypothetical absence of the armed conflict.

In this paper, we examine the causal effect of civil war on Syrian economy, human development and institutional quality by estimating the missing set of counterfactual scenarios using the synthetic control method. Our period of investigation begins in 1996 and ends in 2021, which allows us to exploit a reasonable pre-war outcomes' dynamic to track and reproduce the Syria's economic and institutional development through a convex combination of attributes of countries without armed internal conflict in donor pool of 66 nations. Through the attributes of countries that resemble the latent characteristics of Syria prior to the civil war but have not undergone armed internal conflicts, we are able to estimate the gaps in trajectories of economic growth, human development and institutional quality to plausibly estimate the underlying effect of war.

Our results uncover large and deleterious negative effect of civil war on Syrian economy and society. A broad array of synthetic control estimates invariably suggests that the civil war has caused a large but temporary deviation of economic growth trajectory from its synthetic counterfactual. That said, the civil war has produced a sharp drop in GDP per capita until 2014 that appears to be statistically significant. After 2014, Syrian economy gradually caught up its counterfactual trajectory until the onset of COVID19 pandemic whilst the effect of war on per capita GDP down to the present day has narrowed down substantially. Although a more elaborate analysis would hinge on the specific mechanism at work, the humanitarian and military aid by China, and, to a much lesser extent, Iran and Russia may be an important mechanism behind the rapid postwar recovery.

On the contrary, the effect of civil war on human development is particularly devastating. The evidence based on the series of synthetic control specifications indicates pervasive drops human development capability that is further reinforced by massive decrease in life expectancy, a large increase in infant mortality and a huge drop in population density relative to Syria's synthetic control group. In real terms, our estimates imply that the civil war in Syria has led to at least 5,700 additional infant deaths for each year of the armed conflict which is both unprecedentedly large and also appears to be statistically significant at the conventional levels. Our estimates also uncover some tacit insights into the path of institutional development after the war. In this respect, we show that Syria's institutional quality has deteriorated rampantly towards a more exclusionary, rent-seeking oriented and distortionary institutional environment. Down to the present day, the civil war appears to have caused a large-scale deterioration of the freedom of the expression and pluralism, debilitation of political stability, escalation of political violence against the non-regime forces, rampant drops in government effectiveness and widespread weakening of the rule of law and control of corruption. The estimated institutional quality, economic growth and human development effects appear to be statistically significant and very large. The estimates indicate a rapid recovery of Syrian economy after the first three years of the war coupled with deep losses of human development potential and pervasively more extractive institutional environment tilted in favour of the existing political and business elites that mutually undermines the long-term growth and development potential.

The future research on the effect of Syrian civil war should go a step further. One potentially interesting avenue of research hinges on the subnational effects of civil war by differentiating between regime-controlled areas and the areas under the control of opposition or ISIS insurgency guerrillas, urban and non-urban areas, major and minor cities among several other dimensions to better unravel the heterogeneity of the effects of civil war in greater detail which would greatly improve our understanding the short-term and possible long-term consequences of armed conflicts such as the civil war in Syria.